\documentclass[10pt,a4paper,aps,showpacs,
							amssymb,prl,floatfix,reprint,
							showkeys,superscriptaddress,
							nofootinbib,longbibliography
							]{revtex4-1}

\usepackage[tbtags,intlimits]{amsmath}

\usepackage{slashed}
\usepackage{xcolor}
\usepackage{hyperref}
\usepackage[squaren]{SIunits}

\usepackage[pdftex]{graphicx}
\usepackage{bbold}

\renewcommand{\d}{  {\mathrm d}   }

\newcommand{\Doppler}{\mathfrak D}
\newcommand{\anull}{ a_0 }
\newcommand{\phice}{\phi_\mathrm{0}}
\newcommand{\phispa}{\phi}

\let\vec\boldsymbol
\let\mathbf\boldsymbol
\let\scp\cdot
\renewcommand{\cdot}{\hspace*{-0.1em}\scp\hspace*{-0.1em}}

\begin{document}

\title{Caustic structures in the spectrum of 
x-ray Compton scattering off electrons
driven by a short intense laser pulse}

\author{D.~Seipt}
\email{d.seipt@gsi.de}

\affiliation{Helmholtz-Institut Jena, Fr{\"o}belstieg 3, 07743 Jena, Germany}

\author{A.~Surzhykov}
\affiliation{Helmholtz-Institut Jena, Fr{\"o}belstieg 3, 07743 Jena, Germany}

\author{S.~Fritzsche}
\affiliation{Helmholtz-Institut Jena, Fr{\"o}belstieg 3, 07743 Jena, Germany}
\affiliation{Universit\"at Jena, Institut f\"ur Theoretische Physik, 07743 Jena, Germany}

\author{B.~K\"ampfer}

\affiliation{Helmholtz-Zentrum Dresden-Rossendorf, Institute of Radiation Physics, P.O.~Box 510119, 01314 Dresden, Germany}
\affiliation{TU Dresden, Institut f{\"u}r Theoretische Physik, 01062 Dresden, Germany}


\begin{abstract}
We study the Compton scattering of x-rays off electrons that are driven by a relativistically intense short optical laser pulse.
The frequency spectrum of the laser-assisted Compton radiation
shows a broad plateau in the vicinity of the laser-free Compton line due to a nonlinear mixing
between x-ray and laser photons.
Special emphasis is placed on how the shape of the short assisting laser pulse affects the spectrum
of the scattered x-rays.
In particular, we observe sharp peak structures in the plateau region, whose number and locations
are highly sensitive to the laser pulse shape.
These structures are interpreted as spectral caustics by using a semiclassical analysis of the
laser-assisted QED matrix element.
\end{abstract}

\maketitle



	X-ray free electron lasers (XFELs) help explore
	matter on ultra-short time-scales
	and under extreme conditions.
	Their high x-ray photon flux and short pulse duration of only a few femtoseconds
	allow to record transient processes like chemical reactions in real-time
	\cite{Wernet:Nature2015,Minitti:PRL2015}.
	Moreover, the x-ray scattering off dense plasmas%
	---for instance those plasmas generated by irradiating
	a solid density target with an ultra-intense optical laser pulse \cite{Kodoma:Nature2001}---%
	facilitate the study of ultra-fast
	collective dynamics and plasma instabilities
	\cite{Marklund:RevModPhys2006,Glenzer:RMP2009,Faustlin:PRL2010,Kluge:PhysPlas2014},
	which are important for novel particle acceleration
	\cite{Hatchett:PhysPlas2000,Schwoerer:Nature2006,Metzkes:NJP2014}
	or fusion energy concepts \cite{Kodoma:Nature2001}, for instance.


	Such extreme states of matter can be generated with the help of
	optical lasers that reached already intensities of
	$\unit{10^{22}}{\watt\per\centi\metre^2}$ \cite{Yanovsky:OptEx2008}.
	The interaction of such laser pulses with electrons
	(with charge $e$ and mass $m$) is characterized by the
	laser's normalized amplitude $ \anull = |e|E_L/m\omega_L$,
	where $\omega_L$ and $E_L$ are the frequency and
	amplitude of the laser electric field, respectively.
	Already for $\unit{10^{18}}{\watt\per\centi\metre^2}$ ($\anull \sim 1$),
	the electron's quiver motion reaches relativistic velocities
	and its interaction with the laser's magnetic field leads to a
s	non-linear orbital motion, often denoted as ``figure-8'' \cite{Schappert:PRD1970}.
	At extreme light intensities, $\anull \gg 1$, the electrons
	interact with many photons from the laser field simultaneously and
	one enters the realm of non-perturbative strong-field quantum electrodynamics (QED)
	\cite{Ritus:JSLR1985,Ehlotzky:ProgPhys2009,DiPiazza:RevModPhys2012}.

	High-intensity lasers can also be employed in
	laser-assisted scattering processes
	\cite{Oleinik:JETP1967,Loetstedt:PRL2007,Loetstedt:PRL2008,
	Voitkiv:JPB2003,Muller:JPB2009,Schnez:PRA2007,
	Krajewska:LasPhys2011,Dadi:PRC2012,Szymanowski:PRA1997,
	Li:JPB2004,Meuren:PRL2015,Muller:PRD2014,
	Djiokap:PhysScipta2007,Kanya:PRL2010,
	Boca:CEJP2013,Roshchupkin:Review}, where the presence of the strong low-frequency
	laser field modulates a {hard} QED scattering process.
	This could be, for instance, 
	Compton scattering
	where a hard x-ray (or $\gamma$-ray) photon is scattered off a (quasi-)free electron
	with a frequency change that depends on the scattering angle \cite{Compton:PR1923}.
	The assisting strong low-frequency laser field leads to the formation of side-bands in the
	frequency spectrum close to the laser-free Compton line,
	and already for $\anull \sim 1$ the electron interacts
	with a large number of laser photons \cite{Seipt:PRA2014}.

	In this letter, we present a QED description of laser assisted Compton scattering
	\cite{Oleinik:JETP1968,Gush:PRD1975,
	Akhiezer:JETP1985,Ehlotzky:JPB1989,Nedoreshta:PRA2013,Seipt:PRA2014}
	of an ultra-short pulse of coherent
	x-rays from an XFEL off electrons 
	moving in an intense ($\anull \sim 1$)
	and ultra-short synchronized optical laser pulse 
	\cite{Tavella:NatPhot2011,Hartmann:NatPhoton2014,Schulz:NatCommun2015}.
	We analyze in detail the structures in the frequency spectrum of the scattered x-rays
	with regard to the influence of the specific shape of the assisting laser pulse
	and the ultra-fast laser-driven electron motion.
	By developing a semiclassical picture we
	identify the prominent peaks in the spectrum as
	\textit{spectral caustics} emerging from coalescing stationary phase points where
	the quantum scattering amplitude is formed.
	%

	%
	Caustics are a phenomenon known best from wave propagation.
	They occur when the rays associated to a wave field 
	coalesce on a manifold of lower dimension, creating bright
	zones in the wave field.
	A well known example is the focal spot of a lens:
	All parallel light rays that impinge on the lens coalesce in a single point---the focal point.
	From the mathematical viewpoint, caustics are singularities of differentiable mappings
	\cite{Berry:ProgOpt18,Kravtsov:PhysUspekh1983} and
	also occur in the spectral domain \cite{Raz:NatPhoton2012}.
	The notion of spectral caustics enables us to explain why the plateau region in the
	frequency spectrum is not flat, but has peaks at certain frequencies.


	In our theoretical modeling of laser-assisted Compton scattering
	we describe the incident light of the XFEL ($X$) and assisting laser ($L$)
	as pulsed plane waves with frequencies $\omega_{X,L}$ and durations $T_{X,L}$.
	They copropagate along the $z$-direction, described by the unit four-vector $n^\mu=(1,0,0,1)$,
	with mutually orthogonal linear polarization four-vectors $\varepsilon_{X,L}^\mu$.
	These light pulses scatter off a free electron that has the four-momentum
	$p^\mu$ prior to the interaction.
	We assume the x-rays to be a \textit{weak field} in the sense that
	just one x-ray photon interacts with the electron in a single scattering event \cite{Seipt:PRA2014}.
	The scattered x-ray photon has four-momentum $k'^\mu= \omega' n'^\mu$,
	with frequency $\omega'$ and scattering direction
	$n'^\mu=(1,\,
				\sin\vartheta \cos\varphi ,\, 
				\sin \vartheta \sin \varphi, \,
				\cos \vartheta )$,
	where $\vartheta$ is the scattering angle and
	$\varphi$ denotes the azimuthal angle relative to the laser polarization direction.
	We employ
	units with $\hbar=c=1$
	and the fine structure constant $\alpha=e^2/4\pi$.
	Scalar products between four-vectors are denoted as $x\cdot p = x^0p^0 - \vec x \vec p$.

	Given the above setting, and by employing Volkov states within the framework of strong-field QED in the
	Furry picture \cite{Volkov:1935,Ritus:JSLR1985}, the
	frequency- and angle-differential cross section for laser-assisted Compton
	scattering can be expressed as~\cite{Seipt:PRA2014}
	\begin{align}
	\frac{\d^2\sigma}{\d \omega'\d \Omega} 
			= 
					\frac{\alpha^2 \omega'}{4\pi \omega_L^2 \int_{-\infty}^\infty \d \phi \, g_X^2(\phi)} 
					\frac{\langle|\mathcal  M|^2\rangle}{(n\cdot p)(n\cdot p -n\cdot k')} \,,
		\label{eq:cross_section}
	\end{align}
	where $g_X$ denotes the temporal envelope of the x-ray pulse.
	The squared scattering amplitude,
	{for unpolarized electrons and unobserved polarization of the scattered x-rays},
	is given as a double-integral over the laser phase
	$\phi = \omega_L(t-z)$,
	\begin{multline}
	\langle |\mathcal M|^2 \rangle
				=
			 \int \! \d \phi \d \phi' \, 
			g_X(\phi)g_X(\phi') 
		e^{ i\int_{\phi'}^\phi \! \d \phi'' \, \psi(\phi'',\, \ell) } \\
		\times			\left(
				 \eta - 2 \alpha_X^2 -  \frac{\alpha_X^2}{2} \eta  \: \left[ a_L(\phi) - a_L(\phi')\right]^2
			\right) \,,
	\label{eq:Msquared}
	\end{multline}
	where $a_L(\phi)$ denotes the laser's normalized vector potential
	and we abbreviate $\eta = 2+ \frac{u^2}{1+u}$, $u = \frac{n\cdot k'}{n\cdot p - n\cdot k'}$ 
	and
	$ \alpha_X = \frac{m u}{\omega_X} (\frac{\varepsilon_X\cdot p}{n\cdot p} 
									- \frac{\varepsilon_X\cdot k'}{n\cdot k'})$.
	The phase of the scattering amplitude \eqref{eq:Msquared} is determined by
	\begin{align}
		\psi (\phi,\ell) = (\ell+\varkappa) \frac{n'\cdot v_L(\phi)}{n'\cdot v_0} - \varkappa \,,
	\label{eq:psi}
	\end{align}
	where $\varkappa = \omega_X/\omega_L$ denotes the ratio of the x-ray and assisting laser frequencies.

	The quantity $\ell$ describes the energy transfer from the laser field
	to the scattered x-ray photon
	and determines its frequency $\omega'$ via
	nonlinear x-ray--optical frequency mixing \cite{Glover:Nature2012,Seipt:PRA2014}:
	\begin{align}
	\omega'(\ell) 
	= \frac{(\omega_X + \ell \omega_L) n\cdot p}{ p \cdot n' + (\omega_X + \ell \omega_L) n\cdot n'} \,.
	\label{eq:freq.l}
	\end{align}
	The effective range of $\ell$, and hence also $\omega'$,
	can be quite large even for $\anull < 1$, reaching values of $\ell \sim \anull \varkappa$
	for large frequency ratios $\varkappa$ \cite{Seipt:PRA2014}.

	Expression \eqref{eq:psi} that determines the phase of the scattering amplitude
	depends on the four-velocity of a classical electron moving
	in the laser field $a_L^\mu(\phi)$,
	\begin{align}
	v_L^\mu(\phi) &= v_0^\mu - a_L^\mu (\phi)
										+ n^\mu \frac{a_L(\phi)\cdot v_0}{n\cdot v_0} 
										- n^\mu \frac{a^2_L(\phi)}{2n\cdot v_0} \,,
										\label{eq:velocity} 
	\end{align}
	where $v_0^\mu  = p^\mu/m$ is the electron's four-velocity before the laser pulse arrives.
	In Eq.~\eqref{eq:velocity}, the terms linear in $a_L$ describe the electron's quiver motion
	due to the laser electric field with frequency $\omega_L$.
	The $a_L^2$-term describes the interaction
	with the magnetic field and comprises both a longitudinal $2\omega_L$ oscillation
	and a ponderomotive drift 
	\cite{Schappert:PRD1970,Reiss:PRA2014,Seipt:PRA2015}.
	The superposition of the $\omega_L$ and $2\omega_L$ oscillations
	is often denoted as ``figure-8'' motion.

	\begin{figure}[!t]
		\centering
		\includegraphics[width=0.9\columnwidth]{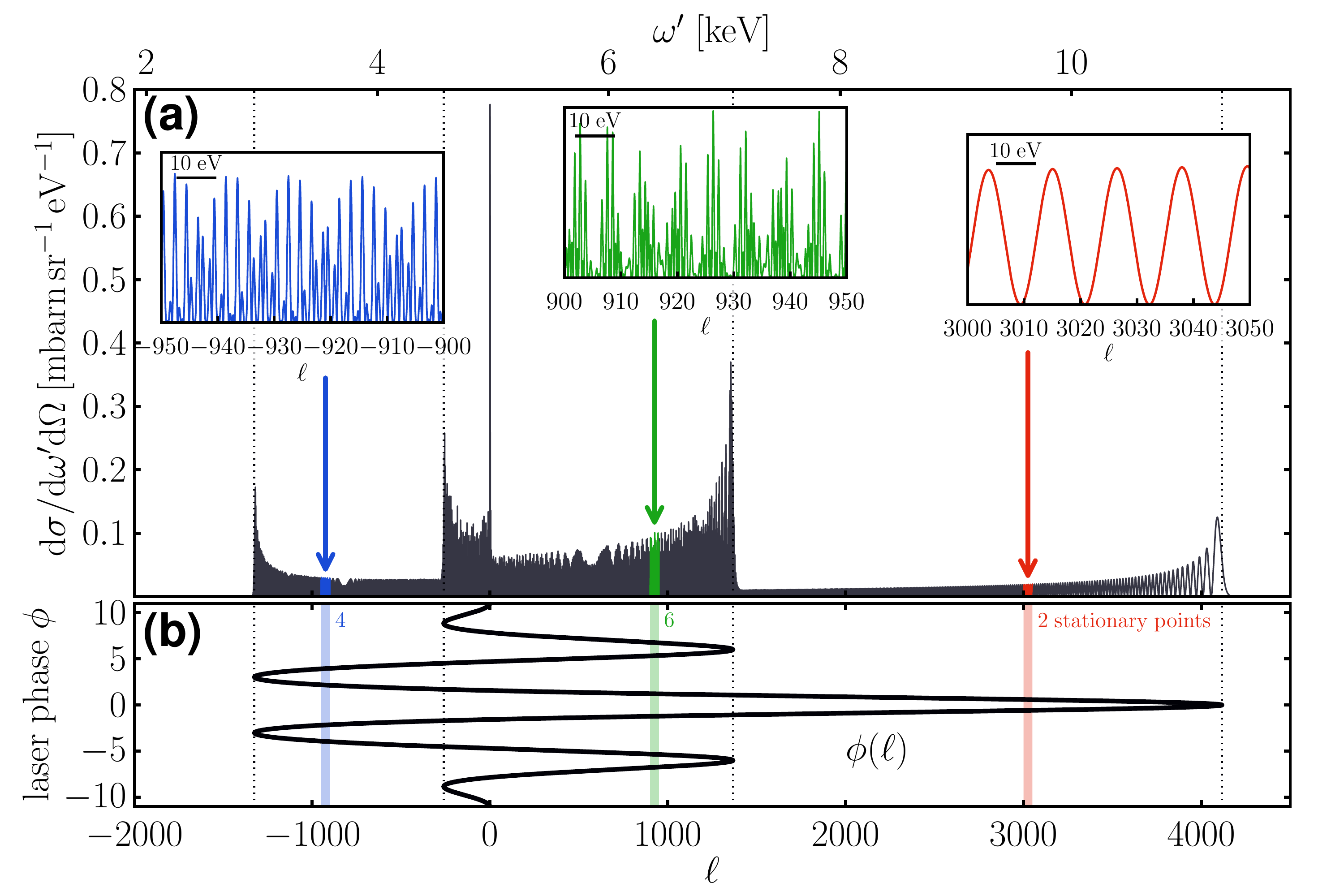}
		\caption{
		Frequency spectrum of x-rays that are
		Compton scattered off laser-driven electrons (a) as a function of
		the scattered x-rays' frequency $\omega'$ (upper axis) and the energy transfer $\ell$ (lower axis).
		The initial x-ray pulse has a frequency of $\omega_X=\unit{5}{\kilo\electronvolt}$
		and a duration of $T_X = \unit{7}{\femto\second}$.
		The assisting laser field that drives the electron-motion
		has peak amplitude $\anull = 1$, frequency $\omega_L=\unit{1.55}{\electronvolt}$
		and pulse duration $T_L= \unit{5}{\femto\second}$.
		The insets magnify the highly-oscillatory structure of the spectrum for three frequency
		ranges.
		Panel (b) depicts the function $\phi(\ell)$ from inverting \eqref{eq:ell.SPA}, 
		which determines the number and positions of the stationary points for a given value of $\ell$.
		Spectral caustics appear at those $\ell$ where the function $\phi(\ell)$ has vertical tangents
		and two stationary points coalesce (vertical dotted lines).
		}
	\label{fig:magnify}
	\end{figure}

	Figure~\ref{fig:magnify} (a) displays
	the frequency spectrum of Compton scattered x-rays,
	Eq.~\eqref{eq:cross_section},
	for a scattering angle of $\vartheta=\unit{45}{\degree}$
	in the plane of the laser polarization, $\varphi=0$.
	For convenience, from now on we work in the rest frame
	of the initial electron, where $v^\mu_0=(1,0,0,0)$.
	We find a narrow large peak of laser-free Compton scattering at $\ell=0$,
	which stems from x-ray photons that scatter
	outside the assisting laser pulse and,
	hence, with no energy exchanged between the electron and the laser field.
	Due to the action of the laser field and the frequency mixing \eqref{eq:freq.l}
	the spectrum has a broad
	structured plateau region between $\unit{3}{\kilo\electronvolt}$ and $\unit{11}{\kilo\electronvolt}$
	with a number prominent large peak structures that we identify as spectral caustics below.
	In the regions between the caustics 
	the spectrum is a highly oscillating function of $\omega'$
	on the scale of $\electronvolt$,
	as can be seen in the insets of Fig.~\ref{fig:magnify} (a).

	To gain an intuitive understanding for the complex peak structure in the spectrum
	in Fig.~\ref{fig:magnify}
	and their relation to the laser-driven electron motion, Eq.~\eqref{eq:velocity},
	it is useful to resort to a semiclassical picture
	by applying a stationary phase analysis \cite{Kaminski:JModOpt2006,Meuren:PRL2015}.
	Because the integrand of \eqref{eq:Msquared}
	is a highly oscillating function of the laser phase $\phi$
	for large frequency ratios $\varkappa=\omega_X/\omega_L \gg1$ and $\anull \sim 1$,
	the scattering amplitude, Eq.~\eqref{eq:Msquared}, is formed mainly at those laser phases
	that fulfill the stationarity condition	$\psi(\phispa,\ell) = 0 $.
	Solving this implicit equation maps the scattering of x-ray photons at
	a certain moment $\phispa$ to one particular value of the energy transfer
	\begin{align}
	\ell(\phispa) = \varkappa \left( \frac{n'\cdot v_0}{n'\cdot v_L(\phispa)} - 1 \right) \,,
	\label{eq:ell.SPA}
	\end{align}
	and, by means of Eq.~\eqref{eq:freq.l}, to one unique frequency $\omega'(\phispa)$.
	The semiclassical analysis of the laser-assisted Compton scattering process facilitates the following
	interpretation:
	The laser-driven electron moves classically according to Eq.~\eqref{eq:velocity},
	up to the laser phase $\phispa$, where the x-ray photon scatters off the electron.
	At the moment of scattering the electron has acquired the velocity $v_L^\mu(\phispa)$
	due to its interaction with the assisting laser field.
	Because the x-ray photon now scatters off a relativistic electron, its frequency is
	Doppler shifted \cite{Compton:PR1923}, and the Doppler shift
	$\Doppler(\phispa) = \frac{n'\cdot v_L(\phispa)}{n\cdot v_0}$
	depends on the angle between the scattering direction $n'$ and
	the instantaneous electron velocity $v_L^\mu(\phispa)$ at the moment of scattering.
	With the help of
	$\Doppler(\phispa)$,
	the instantaneous frequency of the scattered x-rays can be written as
	\begin{align}
	\omega'(\phispa) &= 
	\frac{\Doppler(\phispa) \omega_X }{ 1 + \frac{ n\cdot n'}{n\cdot p} \, \Doppler(\phispa) \omega_X}   \,.
	\label{eq:omega.phi}
	\end{align}
	Thus, in the semiclassical picture
	the broad plateau in the frequency spectrum in Fig.~\ref{fig:magnify} is formed
	because the x-ray photons scatter off accelerated electrons with a variable Doppler factor.

	The structures observed in the plateau-region of the spectrum in Fig.~\ref{fig:magnify}
	can be explained by elaborating further
	the semiclassical picture of the laser-assisted QED scattering process.
	The semiclassical mapping \eqref{eq:ell.SPA},
	relates a moment of scattering $\phispa$ to a \textit{unique energy transfer} $\ell(\phispa)$.
	However, as seen from Fig.~\ref{fig:magnify} (b),
	the inverse function $\phi(\ell)$
	is multiple-valued.
	Thus, the probability
	to observe a scattered x-ray photon with a particular
	frequency $\omega'(\ell)$ is determined by multiple stationary points.
	The contributions to the squared scattering amplitude from 
	different stationary phase points interfere and that leads to the
	highly oscillatory behavior of the frequency spectrum in Fig.~\ref{fig:magnify}.
	For instance, in the region around $\ell \approx 3000$ two stationary points
	contribute to the scattering amplitude,
	leading to a cosine-like oscillation of the spectrum (right inset),
	while around $\ell\approx 900$ (middle inset) a total of $6$ stationary
	points are relevant providing a more complex structure with multiple oscillation periods.

 	All the large sharp peaks in the frequency spectrum are interpreted as spectral caustics:
 	They occur at those values of $\ell$ where two branches of $\phi(\ell)$ merge,
	i.e.~two stationary points coalesce, and the function $\phi(\ell)$ has vertical tangents,
	see Fig.~\ref{fig:magnify}.
	This type singularities of the semiclassical mapping $\phi(\ell)$
	are spectral caustics of the fold-type $A_2$,
	with universal properties \cite{Kravtsov:PhysUspekh1983}.
	The universality allows to estimate the width of the spectral caustic peaks as
	$\Delta\ell \sim (a_0\varkappa)^{2/3}$,
	which is in good agreement with numerical calculations.

	However, how can one understand the 
	existence of the spectral caustic peaks from a physical viewpoint?
	The divergence of $\phi(\ell)$ implies that the caustics are formed at those parts
	of the electron trajectory where the
	Doppler factor $\Doppler(\phi_c)$ becomes stationary, $\dot{\mathfrak D}(\phi_c) = 0$,
	and the scattered x-rays have constant frequency over a long phase region.
	This generates a peak in the spectrum by ``focusing'' the scattered radiation 
	to the spectral caustic peak at $\omega'(\phi_c)$.
	The stationarity of the Doppler factor implies that at the caustic formation phase $\phi_c$
	the four-acceleration of the electron is perpendicular to the scattering direction: 
	$n'\cdot \dot v_L(\phi_c) = 0$.

	\begin{figure}[!t]
		\includegraphics[width=0.49\columnwidth]{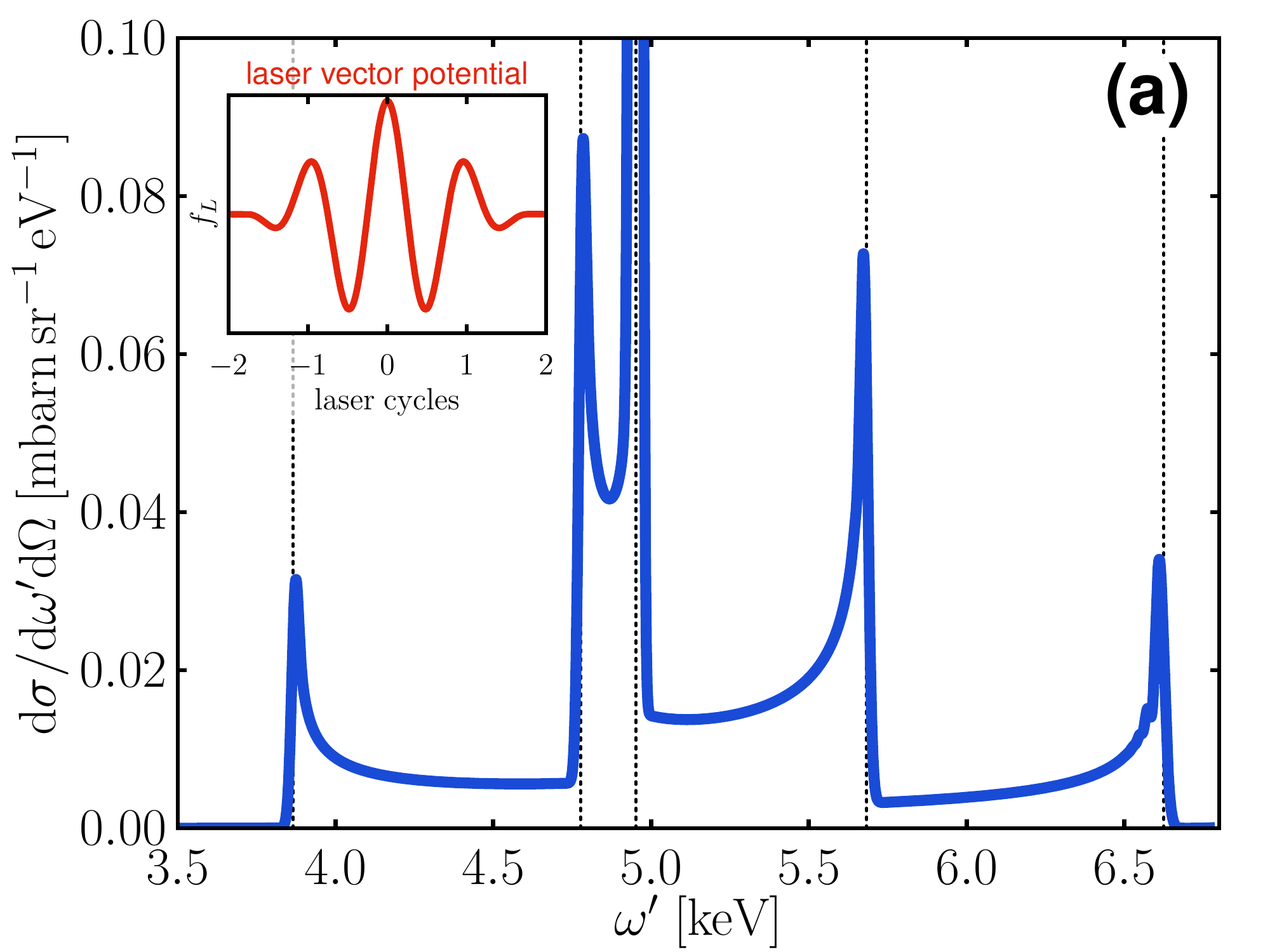}
		\includegraphics[width=0.49\columnwidth]{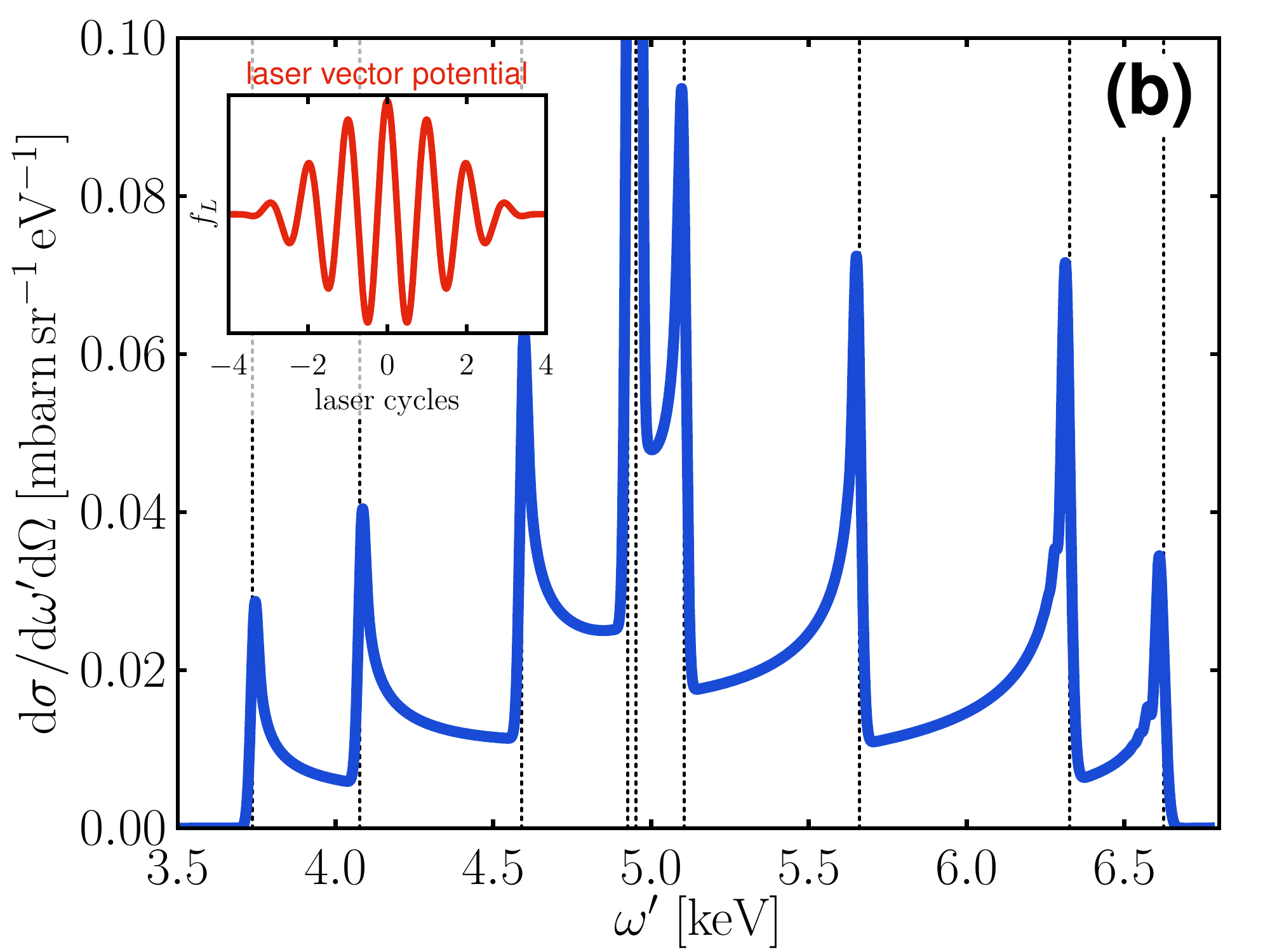}
		\includegraphics[width=0.49\columnwidth]{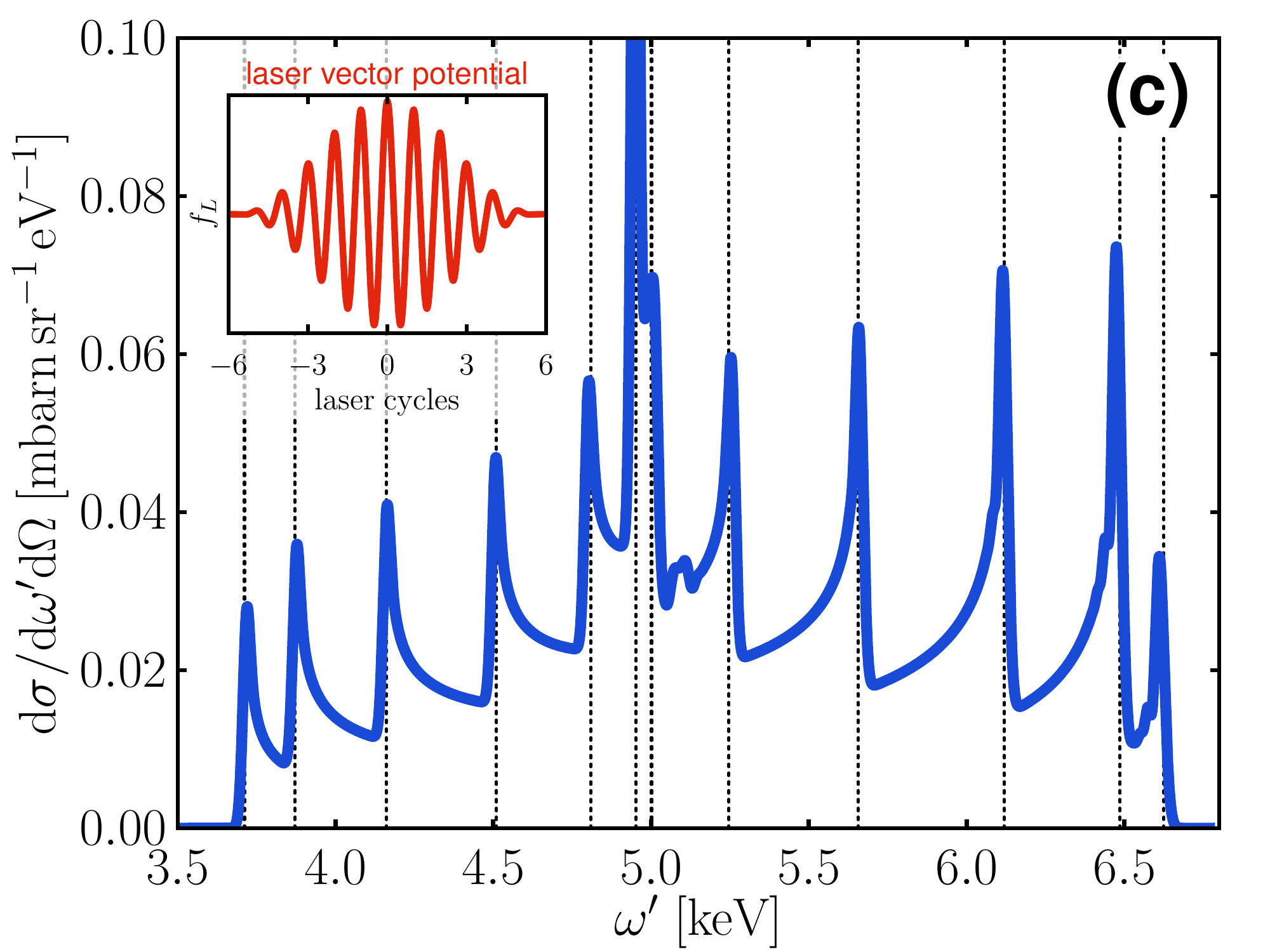}
		\includegraphics[width=0.49\columnwidth]{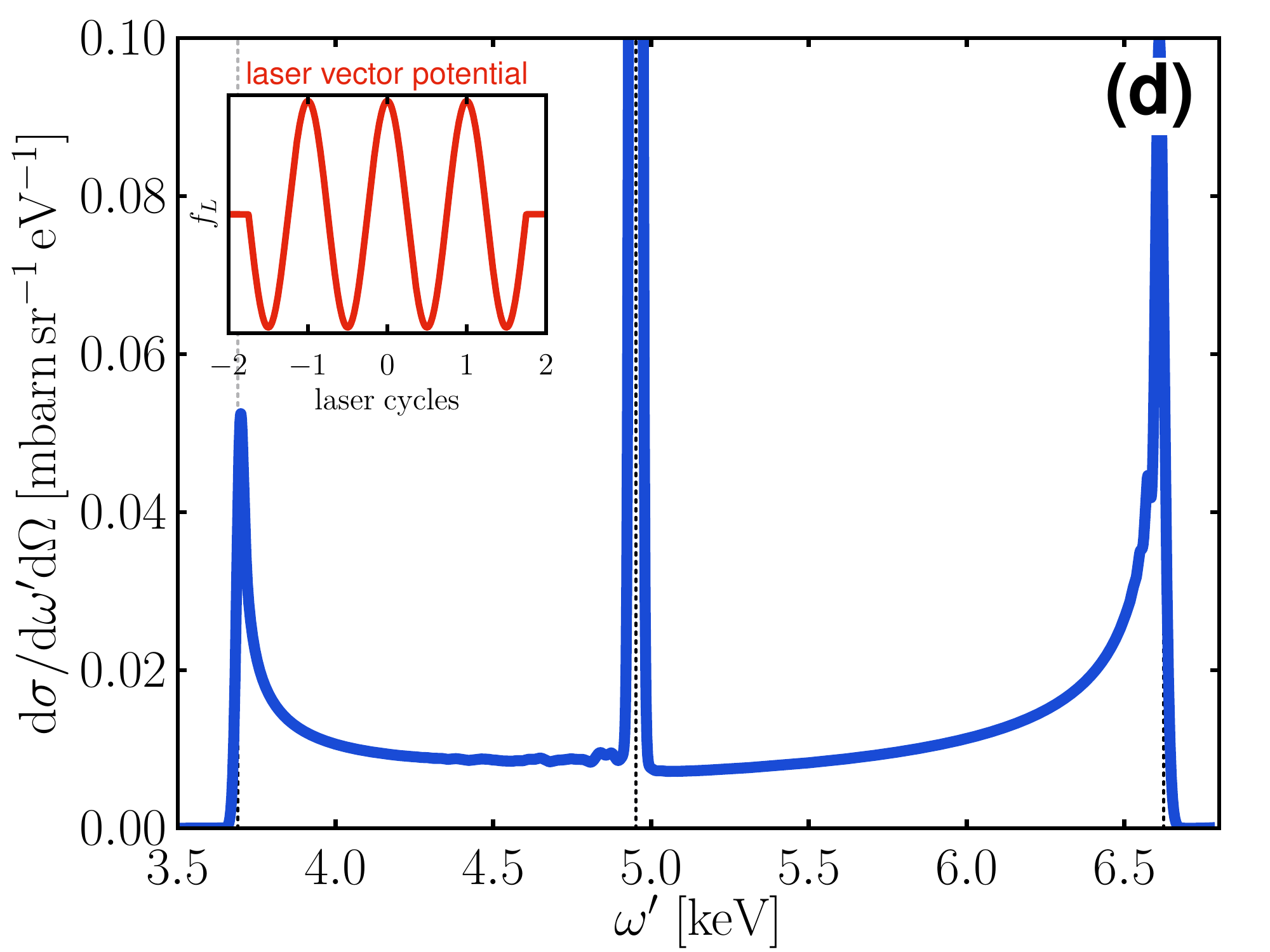}
		\caption{
			The number and locations of the
			spectral caustic peaks in frequency spectrum of the scattered x-rays
			depends on both the shape and duration of the driving optical laser pulse
			(here with peak amplitude $\anull=0.3$).
			The spectra are
			window-averaged (by a Gaussian with resolution $\unit{5}{\electronvolt}$)
			to smooth the fast oscillations.
			Calculations are
			for a squared-cosine pulse
			with pulse durations
			$T_L=\unit{5}{\femto\second}$ (a), 
			$\unit{10}{\femto\second}$ (b) and
			$\unit{15}{\femto\second}$ (c), 
			and for a box shaped pulse with duration 
			$\unit{9.34}{\femto\second}$ (d).
			The vertical lines depict the calculated locations of the spectral caustics.
			The large laser-free Compton peak at $\omega'=\unit{4.95}{\kilo\electronvolt}$
			stems from the long x-ray pulse duration $T_X=\unit{25}{\femto\second} > T_L$.
			}
		\label{fig:pulses}
		\end{figure}

	Let us now calculate, {within the semiclassical picture},
	the locations of the spectral caustic peaks in the frequency spectrum of the scattered x-rays
	and how they depend on the scattering direction $n'$.
	For that, we first have to solve the caustic condition $n'\cdot \dot v_L(\phi_c)=0$
	for the phase $\phi_c$ where the caustics are formed.
	Employing Eq.~\eqref{eq:velocity},
	we can write the caustic condition as
	\begin{align}
	0 	
	 &= \dot f_L(\phi_c)  \left[ \anull f_L(\phi_c) - B(\vartheta,\varphi) \right] \,, \label{eq:nprime.dotu} 
	\end{align}
	with the laser pulse shape $f_L$ and
	$B(\vartheta,\varphi) = \cos \varphi \sin \vartheta \, / ( \cos \vartheta - 1)$.
	Because Eq.~\eqref{eq:nprime.dotu} consists of the product of two terms we
	actually find two different classes of spectral caustics with distinct properties, which 
	we denote as regular and irregular caustics, respectively.

	The regular caustics follow from the solutions of $\dot f_L(\phi_c)=0$.
	For laser pulses $f_L= g_L\cos ( \phi+\phice)$ with a slowly varying envelope $g_L$,
	with $\dot g_L /g_L \ll1$,
	and the carrier envelope phase $\phice$,
	the caustics are formed at the phases $\phi_{c}^{(n)} \approx n\pi-\phice$, $n=0, \pm1, \ldots$.
	For ultra-short pulses, with $\dot g_L /g_L \sim1$,
	the caustic formation phases $\phi_c^{(n)}$ can be obtained numerically.
	The locations of the regular spectral caustics at
	$\ell_\mathrm{reg}^{(n)} = \ell(\phi_c^{(n)}) = \varkappa \xi^{(n)} /(1-\xi^{(n)})$ with
	\begin{multline}
	\xi^{(n)} = (-1)^{n+1} \anull g_L(\phi_{c}^{(n)})  \cos\varphi \sin\vartheta  \\
			- \frac{\anull^2 }{2} g_L^2(\phi_{c}^{(n)}) (1-\cos\vartheta) \,,
	\label{eq:caustic1}
	\end{multline}
	depend on the value of the laser vector potential $a^\mu_L(\phi_{c}^{(n)})$
	at its local extremal points, and, therefore, on the carrier envelope phase $\phice$ and
	on the shape and duration of the pulse, see Fig.~\ref{fig:pulses}.
	The number of different spectral caustic peaks in the panels (a)--(c) grows with
	increasing laser pulse duration $T_L$ for a smooth
	squared-cosine envelope
	$g_L = \cos^2 \frac{ \pi \phi}{2 \omega_L T_L } \, \Theta(\omega_LT_L - |\phi| )$
	with $\Theta$ denoting the step-function.
	For a box-shaped envelope with a constant amplitude
	[Fig.~\ref{fig:pulses} (d)] there are only two regular caustic peaks,
	irrespective of the pulse duration, because of $g_L(\phi_{c}^{(n)} ) = 1$.
	Hence, in order to describe the peaks in the frequency spectrum correctly it is
	essential to exactly take into account the
	shape of the short laser pulse.
	To resolve the individual caustic peaks, their separation should be larger
	larger than their width $\Delta \ell\sim (a_0\varkappa)^{2/3}$.
	This gives the order of magnitude estimate of the optimal laser pulse duration as
	$\omega_LT_L \sim (a_0\varkappa)^{1/3}$.

	\begin{figure}[!th]
	\includegraphics[width=\columnwidth]{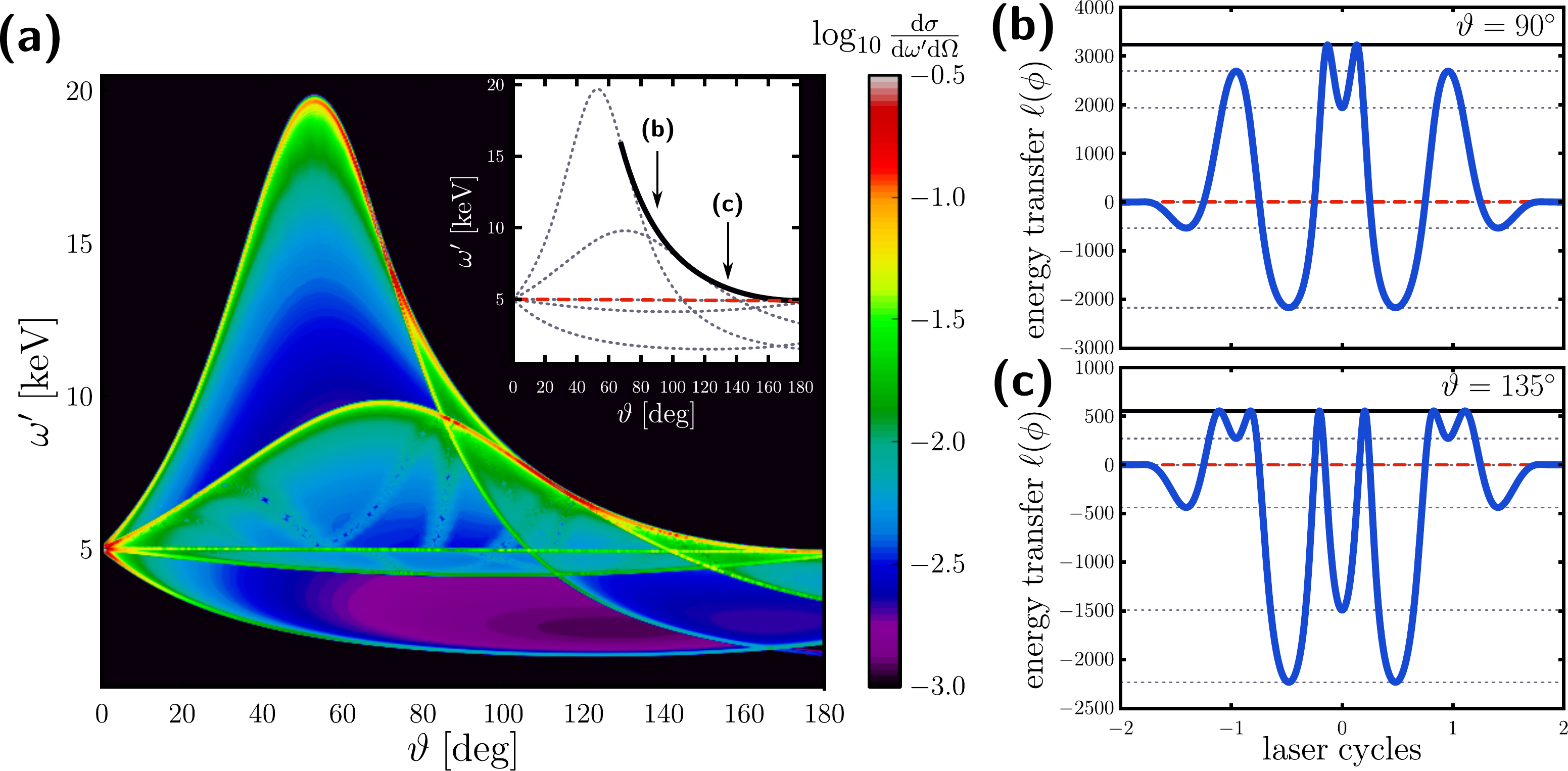}
	\caption{
	The frequency- and angle-differential cross section
	for $\anull =1.5$, windowed by a Gaussian of $\unit{10}{\electronvolt}$ width (a).
	The pattern of the sharp peaks in the spectrum
	qualitatively coincides with the locations of the
	regular (gray dotted curves) and irregular (black solid curve) spectral caustics in the inset.
	The red dashed curve is the laser-free Compton line. Parameters are as in Fig.~\ref{fig:magnify}.
	In (b) and (c) the energy transfer $\ell(\phi)$ oscillates with a frequency $2\omega_L$
	close to the irregular caustics.
	The horizontal lines have the same meaning as in the inset of panel (a).
	}
	\label{fig:2d}
	\end{figure}

	The second type of caustics---the irregular caustics---%
	occur where $f_L(\phi_c) = B(\vartheta,\varphi)/\anull$ admits at least one real solution for $\phi_c$.
	In stark contrast to the regular caustics discussed above the location of the
	irregular caustic peak at
	$\ell_\mathrm{irr} = \varkappa \zeta/(1-\zeta)$, with
	$\zeta = (1 - \cos\vartheta )B^2(\vartheta,\varphi)/2$,
	is independent of the laser pulse parameters.
	The irregular caustics are related to the longitudinal non-linear motion of the electrons due to the
	$a_L^2$-term in the classical electron velocity, Eq.~\eqref{eq:velocity}.
	Therefore, they occur only for large enough scattering angles,
	e.g.~$\vartheta>\unit{65}{\degree}$ in Fig.~\ref{fig:2d},
	where one probes dominantly longitudinal components of the electron velocity \eqref{eq:velocity}.
	This is related to the forward-backward asymmetry seen in Fig.~\ref{fig:2d} (a).
	The Doppler up-shift of the x-ray frequency for backward-scattering
	is limited by the longitudinal ponderomotive drift of the electron
	away from the observer.
	The existence of the irregular caustic peak in the spectrum signals
	the non-linear relativistic motion of the electrons, which comprises both the longitudinal
	ponderomotive drift and $2\omega_L$ oscillations.
	In fact, the semiclassical mapping $\ell(\phi)$,
	Eq.~\eqref{eq:ell.SPA}, shows distinct $2\omega_L$-oscillations
	wherever the irregular caustics exist, see Fig.~\ref{fig:2d} (b,c).


	Experiments on laser-assisted Compton scattering to verify the spectral caustic peaks in the spectrum
	could be done, e.g.~at the future HIBEF beamline at the European XFEL \cite{Cowan:HIBEF2013}
	or the LCLS,
	with an XFEL pulse and an ultra-short intense optical
	laser pulse impinging simultaneously on an electron beam
	\cite{Seipt:PRA2014}.

	Due to the sensitivity of the locations of the spectral caustic on the laser pulse shape,
	the laser-assisted Compton scattering could give promise to measure
	the properties of relativistically intense laser pulses,
	in particular its peak amplitude $a_0\gtrsim 1$, pulse duration and carrier envelope phase,
	for intensities exceeding the capabilities of
	atom-based schemes as stereo-ATI \cite{Paulus:Nature2001,Wittmann:NatPhys2009}.

	The laser-assisted Compton scattering may be used to observe the dynamics of
	laser-driven electrons in more general situations, where the electrons are also subject to
	forces other than the laser field.
	The three-dimensional electron motion could
	be accessed by observing tomographically
	the frequency spectrum of Compton scattered x-rays for different scattering directions $n'$.
	For instance, this might be useful to investigate the
	complex laser-driven electron dynamics at the surface of a dense plasma,
	and could help to better understand
	the collisionless absorption of laser energy
	\cite{Brunel:PRL1987,Mulser:PhysPlas2012},
	and its implications for plasma-based particle acceleration \cite{Veltcheva:PRL2012}.
 
	In summary, we study for the first time the details of the frequency spectrum of x-rays that are Compton
	scattered off an electron under the action of an intense \textit{ultra-short} laser pulse.
	The rich pattern of large sharp
	peaks in the frequency spectrum is a novel feature that is related to the ultra-short
	duration of the assisting laser pulse, and is explained by means of a semiclassical picture
	as spectral caustics with universal properties.
	The notion of spectral caustics is a general one and could help
	understand
	also	
	other laser-assisted scattering processes in ultra-short laser pulses.

%


\end{document}